\documentstyle[11pt,aas2pp4,psfig]{article}   

\begin{document}

\newcommand{\Mo}{\mbox{$\rm M_\odot$}}
\newcommand{\Lo}{\mbox{$\rm L_\odot$}}
\newcommand{\mic}{\mbox{$\rm \mu m$}}
\newcommand{\ivol}{\mbox{$\rm cm^{-3}$}}
\newcommand{\isup}{\mbox{$\rm cm^{-2}$}}
\newcommand{\isec}{\mbox{s$^{-1}$}}
\newcommand{\Av}{\mbox{$A_{\rm V}$}}
\newcommand{\Ne}{\mbox{$N_{\rm e}$}}
\newcommand{\Te}{\mbox{$T_{\rm e}$}}
\newcommand{\ten}[1]{\mbox{$10^{#1}$}}
\newcommand{\xten}[1]{\mbox{$\times 10^{#1}$}}
\newcommand{\wl}{\mbox{$\lambda$}}
\newcommand{\forb}[2]{\mbox{$[{\rm #1\, #2}]$}}
\newcommand{\ha}{\mbox{H$\alpha$}}
\newcommand{\hb}{\mbox{H$\beta$}}
\newcommand{\hg}{\mbox{H$\gamma$}}
\newcommand{\he}{\mbox{H$\epsilon$}}
\newcommand{\hd}{\mbox{H$\delta$}}
\newcommand{\nehe}{\neiii+\he}
\newcommand{\heii}{\mbox{He\,II}}
\newcommand{\oiii}{\forb{O}{III}}
\newcommand{\neiii}{\forb{Ne}{III}}
\newcommand{\ariv}{\forb{Ar}{IV}}
\newcommand{\oii}{\forb{O}{II}}
\newcommand{\oi}{\forb{O}{I}}
\newcommand{\nii}{\forb{N}{II}}
\newcommand{\sii}{\forb{S}{II}}
\newcommand{\fevii}{\forb{Fe}{VII}}

\lefthead{Axon et al.} 
\righthead{Jet driven motions in NGC 1068}

\title{Jet driven motions in the\\
Narrow Line Region of NGC 1068
\footnote{Based on
observations with the NASA/ESA Hubble Space Telescope, obtained at the
Space Telescope Science Institute, which is operated by AURA, Inc.,
under NASA contract NAS 5-26555 and by STScI grant GO-3594.01-91A}}

\slugcomment{Version as of \today}

\author{D.J. Axon\altaffilmark{2,3}, A. Marconi\altaffilmark{4,5}}

\affil{Space Telescope Science Institute \\
       3700 San Martin Drive\\
       Baltimore, MD 21218}
\author{A. Capetti}
\affil{Osservatorio Astronomico di Torino\\
Strada Osservatorio 25\\
40025, Pino Torinese, Italy}
\author{F.D. Macchetto\altaffilmark{2}, E. Schreier}
\affil{Space Telescope Science Institute}
\author{A. Robinson}
\affil{Division of Physics and Astronomy, \\
 Department of Physical Sciences, \\
 University of Hertfordshire,\\
 College Lane, Hatfield, Herts AL10 9AB,UK}  
\altaffiltext{2}{Affiliated to the Astrophysics Division, Space Science
Department, ESA}
\altaffiltext{3}{On leave from the University of Manchester}
\altaffiltext{4}{Dipartimento di Astronomia e Scienza dello Spazio,
Universit\`a di Firenze, Largo E. Fermi 5, I--50125, Italy}
\altaffiltext{5}{Osservatorio Astrofisico di Arcetri,
Largo E. Fermi 5, I--50125, Italy}

\authoremail{axon@stsci.edu}

\begin{abstract}
We have obtained HST FOC f/48 long--slit spectroscopy of the inner 4
arcseconds of the Narrow Line Region of NGC 1068 between 3500-5400\AA\
with a spectral resolution of 1.78\AA/pixel. At a spatial scale of
0\farcs0287 per pixel these data provide an order of magnitude
improvement in resolution over previous ground based spectra and allow
us to trace the interaction between the radio jet and the gas in the
NLR.  Our results show that, within $\pm 0\farcs5$  of the radio-jet the
emission lines are kinematically disturbed and split into two components whose velocity separation
is 1500 km \isec.  The filaments associated with the radio lobe also show a redshifted kinematic disturbance of the order of 300 km \isec which probably is a consequence of the expansion of the radio plasma.

Furthermore, the material
enveloping the radio-jet is in a much higher ionization state than that
of the surrounding NLR gas.  The highest excitation is coincident with
the jet axis where emission in the coronal line of [FeVII] $\lambda
3769$\AA\ is detected and the \heii$\lambda$4686 \AA\ is strong
but where [OII] $\lambda$3727 \AA\ is depressed. 
This large localized increase in ionization on the jet axis is accompanied by the presence of an excess continuum. Because the electron density is substantially larger in the jet compared to the surrounding NLR, these results can only be explained if 
there is a more intense ionizing continuum associated with the jet. This can be accomplished in a variety of ways which include an intrinsically anisotropic nuclear radiation field, a reduced gas covering factor
or the presence of a local ionization source.

The morphology, kinematics and, possibly, the ionization structure
of the NLR in the vicinity of the
jet of NGC 1068 are a direct consequence of the interaction with
the radio outflow.

\end{abstract} 


\section{Introduction}

Extensive HST
emission line imagery of the Narrow Line Regions (NLR) of Seyfert galaxies
with linear radio sources has shown that the morphology of the NLR is
directly related to that of the radio emission.  Seyfert galaxies with a
lobe-like radio morphology (e.g.  Mrk 573, Mrk 78, NGC 3393, IRAS 0421+045
and IRAS 1105-035) have bow-shock shaped emission line regions
(\cite{bow94,axhst,mrk573}), while those with a
jet-like radio structure (e.g.  Mrk 3, Mrk 348, Mrk 6, Mrk1066) have jet-like
emission line structures (\cite{bow95,mrk3,mrk6}).
These results have provided compelling evidence for strong
dynamical interactions between the emission line gas and radio-emitting
ejecta on scales of $\leq1$\,kpc.

The physical foundation of this picture was
first developed by Taylor, Dyson and Axon (1992) who constructed fast bow
shock models for the interaction between radio ejecta and the ISM 
which included the effects of photo-ionization from the nucleus. 
Dopita and Sutherland (\cite{dop96a,dop96b}) have emphasized
that the hot shocked gas could itself make an important local contribution to the ionizing continuum  (becoming what they termed an {\sl auto-ionizing} 
shock).

NGC 1068 is one of the closest Seyfert galaxies and it harbours a bright
radio source, with a prominent radio-jet (\cite{mux96}) terminating in
an extended radio-lobe (\cite{wil83}). Ground-based kinematic
studies of the NLR have been carried out previously by a number of authors
(\cite{al83,bal87,me86,cec90,ung92,arr96}) in [OIII]
$\lambda 5007$.
In particular two broad components straddling a
narrow central component in the NE radio lobe were identified.
The narrow component appears to
follow the expected rotation curve of the galaxy and is emitted by
undisturbed gas in the disk.  At larger radii this triple structure
disappears.

The complexity of the velocity field observed in the inner region is
almost certainly due to the spatial confusion caused by
projection and seeing effects
which mixes kinematic components from different regions 
along the line of sight.  In particular the
velocity structure created by the expansion of the hot gas associated with
the lobe is projected onto the region interacting
with the jet which 
can be spatially resolved only with HST. In order to obtain a more
detailed picture of the relationship between the velocity field and the
ionization conditions in
the NLR gas and the interaction of the gas with the radio jet we have therefore
obtained new long--slit spectroscopic observations  with the Faint
Object Camera f/48 spectrograph on board the HST.
Our results allow us to
resolve the spatial confusion and provide evidence that shocks
created by the jet play a key role in forming and, possibly, ionizing
the NLR of NGC 1068.
Throughout this paper we will adopt a distance to NGC 1068
of 14.4 Mpc (\cite{tully})  where 1\arcsec\ corresponds to 72 pc.

\begin{figure*}
\centerline{\psfig{figure=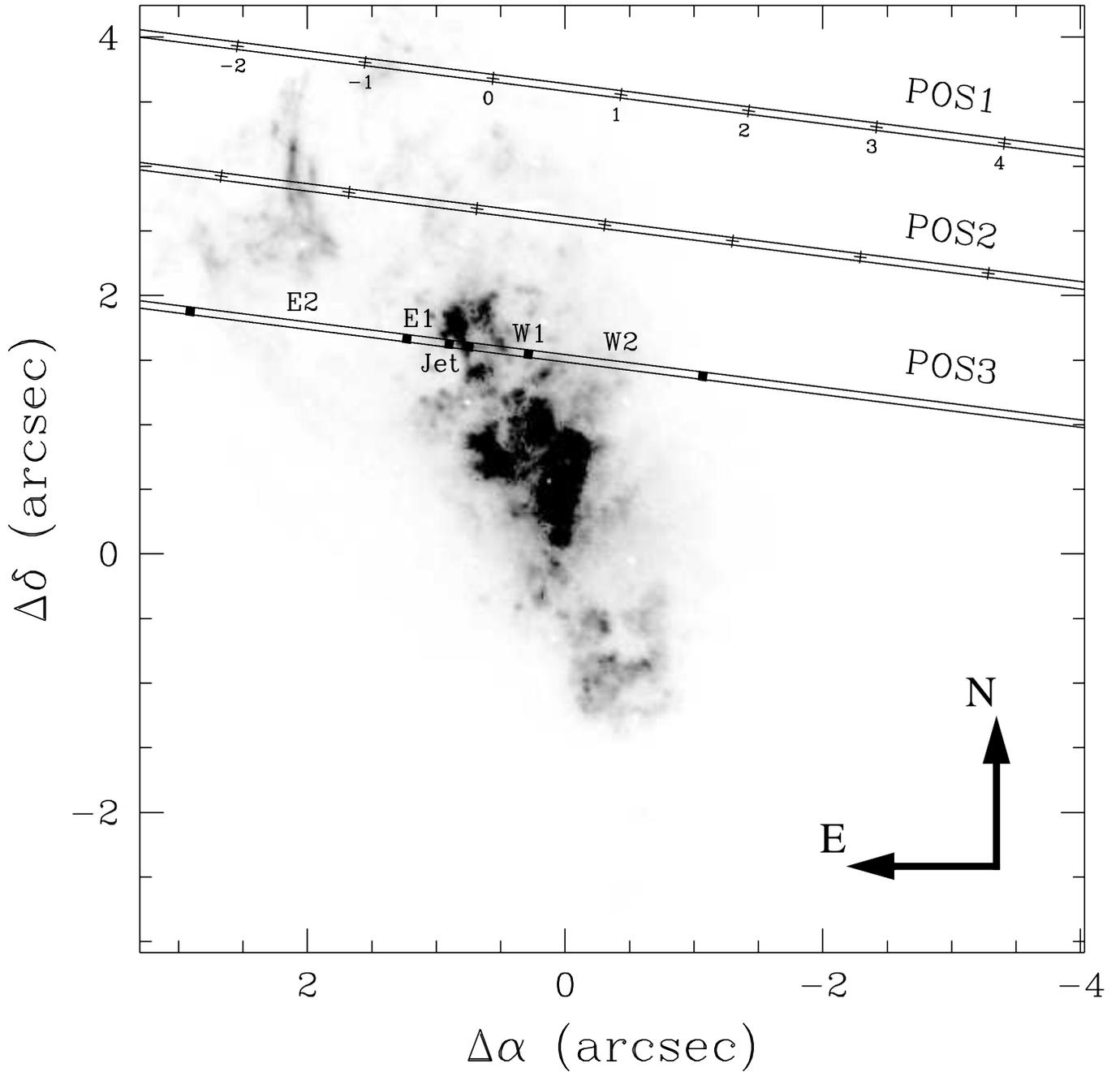,angle=-90,height=\linewidth}}
\caption{\label{fig:slitpos} 
The slit positions used in our study overlayed on a gray scale map of
the \oiii\ image of Macchetto et al.~1994. The origin of the spatial axis
on the borders of the image is centered on the position of the
hidden nucleus (Capetti et al.~1995c). The spatial coordinates used along
the slit are marked as crosses at POS1 and POS2.
The zero point of this axis lies on cloud G in POS3. The
filaments associated with the radio lobe intersect the slit at POS2 at
location  -1.5 arcseconds  and POS1 at -1 arcseconds.
The filled squares at POS3 indicate the limits of the contiguous regions
where we extracted the spectra shown in Fig. 3.}
\end{figure*}

\section{Observations and Data Reduction}
The Narrow Line Region of NGC 1068 was observed using the FOC
f/48 long--slit spectrograph 
on October 18$^{\rm th}$, 1996 at resolutions of  1.78\AA\ 
and 0\farcs0287 per pixel along the dispersion and slit directions,
respectively.
The F305LP filter was used to isolate the first order
spectrum which covers the 3650--5470 \AA\ region and therefore includes
the \oii\wl 3727, \hb\wl 4861 and \oiii\wl\wl 4959, 5007 \AA\  emission lines.
The slit, 0\farcs063x13\farcs5 in size, was placed at a position angle
of 83$^\circ$ and spectra with exposure times
of 627 seconds were taken in the 1024x512 non--zoomed mode
at 6 locations separated by 1\arcsec. Unfortunately, due to a guide star re-acquisition failure, only
3 of the slit locations yielded usable spectra which we identify as POS1,
POS2 and POS3 respectively.

The data reduction follows the procedure described
in detail by \cite{m87}. Flux calibration was
performed using the observations of the spectrophotometric
standard star LDS749b.

To accurately determine the location of the slits we compared the
surface brightness profile derived from the FOC, f/96, F501N HST image of
\cite{ero} with that measured from the spectra at the three slit positions.
The best match is displayed in Fig. \ref{fig:slitpos}
and is accurate to within half a slit width ($\simeq0\farcs03$).

\begin{figure*}
\centerline{\psfig{figure=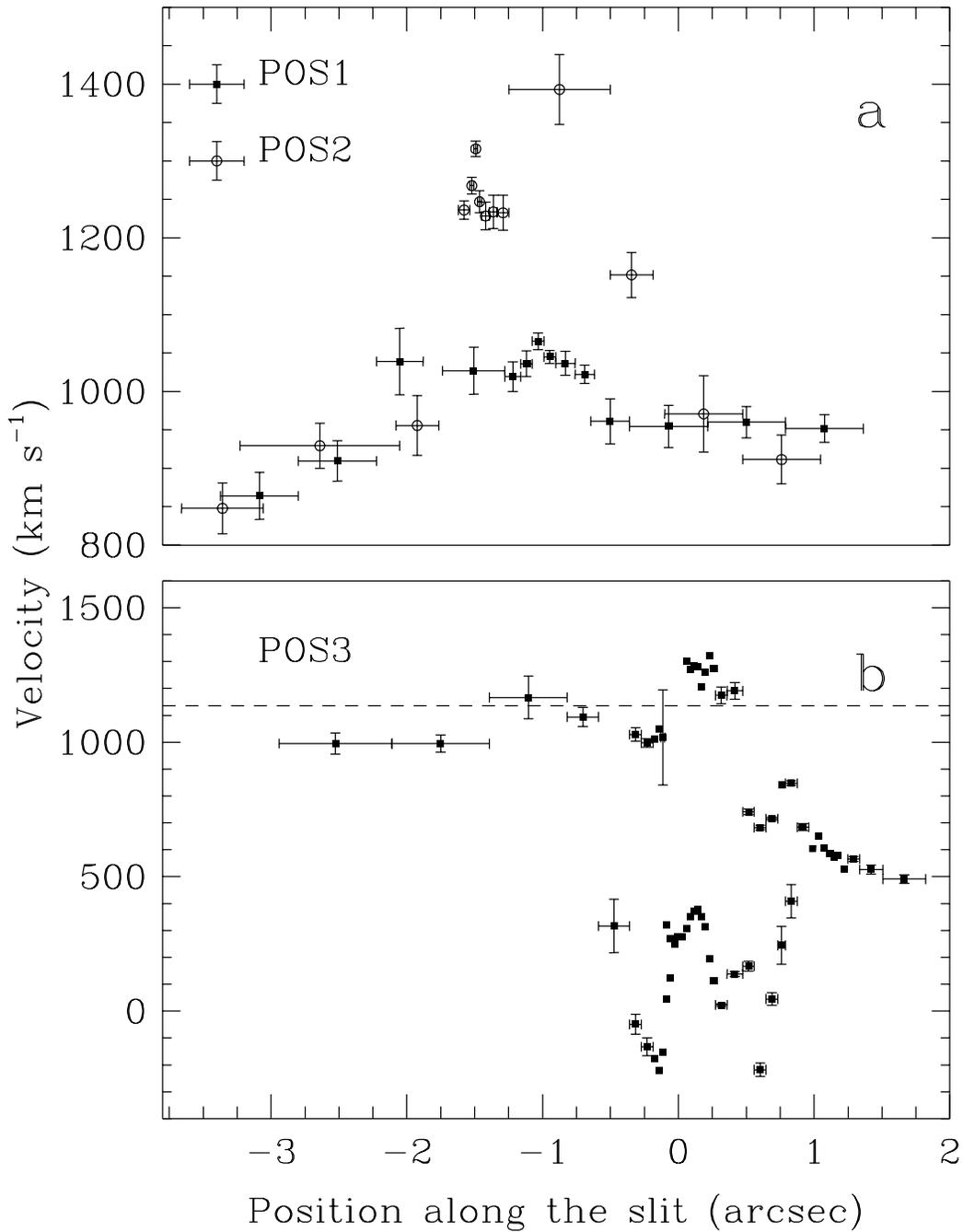,width=0.8\linewidth}}
\caption{\label{fig:vel} 
Velocity fields measured at the three slit positions.
Lower panel:
Around the jet axis, at POS3,
the gas is strongly kinematically perturbed and shows split lines with
components separated by $\sim$ 1500 km \isec.
The dashed line marks the systemic velocity of NGC 1068
(Baan \& Haschick 1983).
Upper panel:
at POS2 and POS1 there are also large velocity perturbation at 
the location of the emission line filaments.}
\end{figure*}

The slit locations POS1 and POS2 cut across the emission line filament system associated with the Northern radio lobe (\cite{cap97}). While the
slit at POS3 (see Fig. \ref{fig:slitpos}) traverses cloud G 
(\cite{eva91})
and the parallel feature to the West which form a funnel-like structure where the jet meets the radio lobe (see \cite{gal96}).

\section{\label{sec:results} Results}

\begin{figure*}
\centerline{\psfig{figure=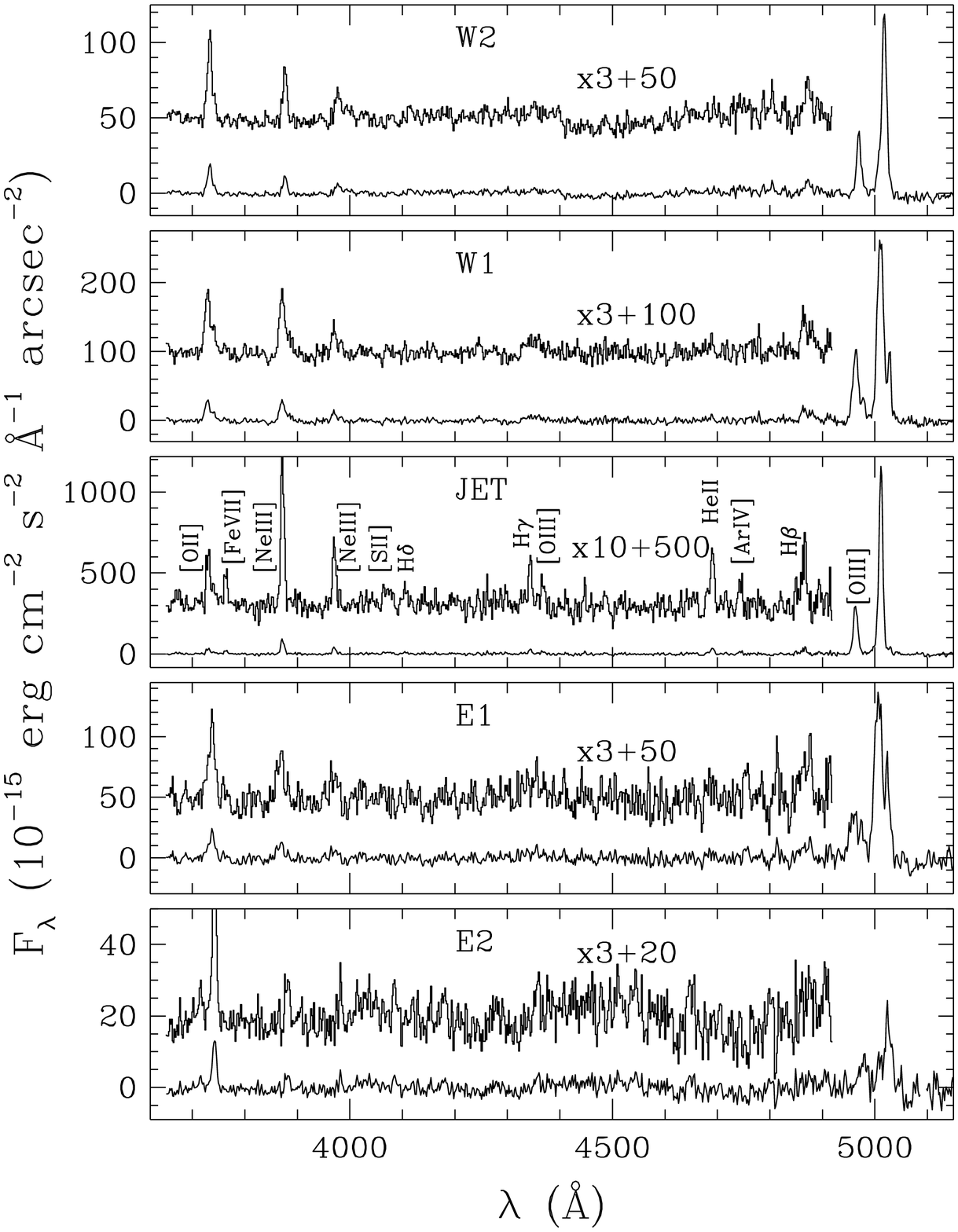,width=0.8\linewidth}}
\caption{\label{fig:spectra} Spectra extracted at the positions
indicated in Fig. 1.
The spectrum extracted in a 0\farcs14$\times$0\farcs063 aperture around
cloud G, labeled "Jet", shows the high
excitation gas on the jet with \fevii, \ariv\ and \heii\  emission lines.}
\end{figure*}
The velocities derived from the  \oiii\ and \oii\ lines by fitting
gaussian line profiles are plotted in  Fig. \ref{fig:vel}. Around the
jet axis, at POS3 (Fig. \ref{fig:vel}b),
the lines are split into two velocity systems
separated by $\sim$ 1500 km \isec.  Outside this region the gas motions
are quiescent but with a gradient of +500 km \isec\ from West to East.
Notice that the velocity splitting is not symmetric about the systemic
velocity, but is much larger on the blue-shifted side. The blueshifted
component is also substantially brighter and has a larger velocity
dispersion than the redshifted component. At POS2
(Fig. \ref{fig:vel}a) the velocity field is
relatively flat, except at the filaments at slit coordinates
$\sim$ -1\arcsec\ -- -2\arcsec, where there is a large redshifted velocity perturbation
of amplitude $\sim$ 300 km \isec, at a similar radial velocity to that of
the redshifted component of the split line region in POS3. POS1
(Fig. \ref{fig:vel}a) shows a similar pattern in velocity, 
but with a reduced amplitude in the perturbed region.

\begin{figure*}
\centerline{\psfig{figure=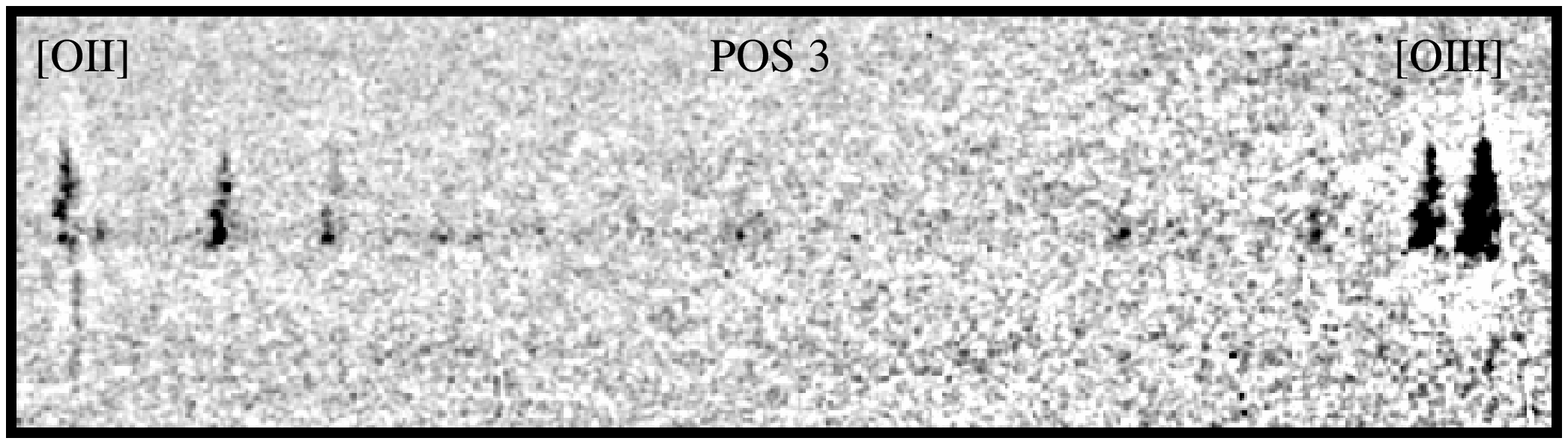,angle=-90,width=\linewidth}}
\caption{\label{fig:new} Grey scale image of the spectrum at POS3 showing the excess continuum associated with the radio jet.}
\end{figure*}

Turning to the emission line strengths, in Fig. \ref{fig:spectra} we show plots of representative spectra from the five regions identified in
 Fig. \ref{fig:slitpos}.
On the jet axis (labeled as ``Jet'') the gas is highly ionized;
\fevii\wl 3759\AA\ and \ariv\
are clearly visible and the \heii\wl4686 is as high as 0.7 the strength
of \hb, while the \oii\ emission appears to be depressed.
This increase in excitation is accompanied by the presence of an excess
 local continuum, as shown in Fig. \ref{fig:new}.

\begin{figure*}
\centerline{\psfig{figure=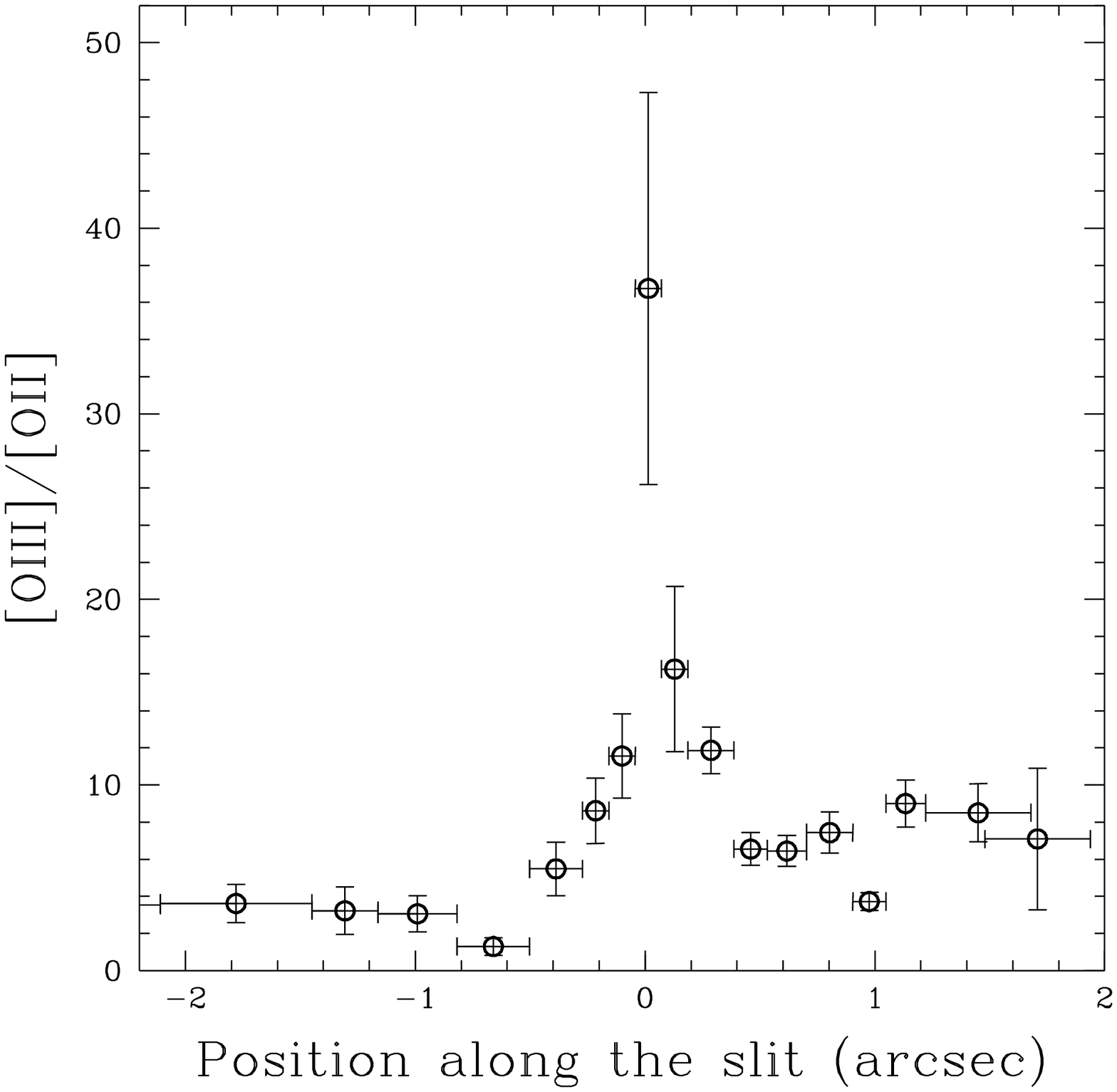,width=0.6\linewidth}}
\caption{\label{fig:ratio} A plot of the \oiii/\oii\ ratio 
as a function of position along the slit at POS 3. The ratio reaches a maximum on the jet axis where it is $\sim$ 5 times larger than
that in the surrounding regions.}
\end{figure*}

Outside the jet the excitation conditions are much lower, with \oii\
being relatively stronger, particularly to the E as shown in Fig.
\ref{fig:new}.  The variation of the \oiii/\oii\ ratio along the slit,
plotted in Fig. \ref{fig:ratio}, shows that there is a factor $\simeq$5
difference between the ratio on the jet axis and outside. The electron
temperature measured from the \oiii\wl\wl 4363/5007 ratio give
$\Te=(1.15\pm0.15)\xten{4}$ K. A lower limit for the
electron density can be obtained from the \ariv\wl\wl 4710/4740 ratio
and yields $\Ne\ge\ten{4.5}\ivol$.
At similar distances from the nucleus, but off the jet,
ground based density measurements from \sii\wl6716,6731 ratio, obtained
from our unpublished spectroscopy, indicate electron densities of a few
100 \ivol.  Thus the increase in excitation on the jet
occurs despite an increase in density.

\section{\label{sec:discussion} Discussion}

The morphological
connection between the optical and radio emission in the  region of the
jet has been discussed in detail by Gallimore et al. (1996) and by
Capetti, Macchetto and Lattanzi (1997).  They showed that there is a
clear anti-correlation between radio and optical emission, with the
brightest line emitting knots surrounding the radio-jet.  In the
jet-cloud interaction picture, the radio jet clears a channel through the cool
ISM, leaving hot ionized gas in its wake 
and the line emission is highly enhanced along the edges of
the radio-jet where the gas is compressed.

The overall velocity structure and ionization conditions of the gas
associated with the jet of NGC1068 can be understood in terms of this
picture of an expanding and cooling cocoon around the jet which 
creates very fast shocks with at least $V \simeq 700$ km \isec\
(even when mass-loading of the expanding cocoon is taken into account).

The viewing angle is such that to the NE the lobe is projected on to the
disk of NGC1068, so that the jet is seen emerging towards us out of the
disk. The situation is very similar to that seen previously on the jet
of 3C120 (\cite{3c120}), even down to the clumpy structure of the gas,
which almost certainly arises because of the onset of instabilities
caused by the mixing of hot and cold material.  This morphology is again
readily understood if it is due to the interaction between the jet and
the surrounding medium. A viable explanation for the observed asymmetry
of the velocity splitting is that it is due to density stratification of
the disk, that is the expanding hot bubble is able to expand more
rapidly along the density gradient away from the disk plane, than
towards it. Further weight to this argument is provided by the systematic blue-shift of coronal lines with respect to the lower
excitation lines (\cite{mar96}). This implies that the coronal lines comes from a different kinematic component which we would ascribe to the hot cooling gas in the cocoon.

The spectral diagnostics obtained from our long-slit spectra indicate that
the kinematically disturbed line-emitting gas associated with the radio
jet has both a higher density and a higher excitation level than the
surrounding NLR. A similar result, based
on narrow band imaging, has been reported recently by \cite{cap97}.
The density enhancement is a factor $>$ 100 if the
ground-based measurements of the [SII]\wl\wl 6716,6731 ratio can be taken as
typical of the regions extending away from the jet axis.  On the other
hand, the variations in the [OIII]5007/[OII]3727 ratio along the POS 3
slit imply a factor 5 enhancement in the ionization parameter (U) across
the jet axis (since this ratio is approximately proportional to U; Penston
et al 1989).  The enhancements in both density {\sl and} ionization parameter
imply that the local ionizing radiation is a factor $\sim$ 500 higher at the
location of the jet in POS 3, than it is elsewhere. The exceptionally
strong [Fe VII] and HeII lines suggest that matter-bounded clouds might
contribute significantly to the line emission near the jet. In this case,
we will be overestimating the jump in U since low ionization lines such as
[OII] are strongly suppressed in such clouds. Nevertheless, this is
unlikely to explain the whole of the inferred increase in the ionizing flux.
Either a local ionization source is present which dominates the AGN
radiation field near the jet axis, or the AGN radiation field is itself
highly anisotropic. 

The most important local ionization source is likely to be ionizing
radiation (free-free and high ionization line emission) from gas heated
and ionized in fast shocks driven by the radio jet (\cite{binette}; 
\cite{sut93}; Dopita \& Sutherland 1996a,b).  The strong [Fe VII]
emission and the excess continuum observed on the jet axis
may be a clue that the coronal gas expected in this picture is
indeed present.
Alternatively, the on-axis increase in the ionizing flux
could be attributed to intrinsic anisotropy of the nuclear continuum.  In
this case, the size of the enhancement in flux, and the fact that it is
highly localized, require a continuum beaming factor comparable with
Doppler boosting in a moderately relativistic jet (gamma ~ a few). 
This idea can be ruled out because
the region we have studied occurs after a significant bend of the jet
and therefore one would expect to see higher excitation along the
orginal jet  direction, which is not observed (cf. \cite{cap97}).

Perhaps a more likely explanation is that the
continuum anisotropy is largely produced by azimuthal variations in the
optical depth to ionizing photons.  In this picture, the radio jet sweeps
out a channel as it passes through the NLR with the result that ionizing
radiation from the central source suffers much less attenuation along its
axis than in other directions. 

Our results on POS1 and POS2 can also be interpreted in terms of the interaction between the radio plasma and the ambient gas, but in this case it is the expanding lobe material which is driving the motion. 
From our spectroscopic data we can say little on the ionization conditions of these filaments because only the [O III] lines are detected.

\section{Conclusions}
  
We have presented new HST FOC f/48 long--slit spectra of the central 4
arcseconds of the NLR of NGC 1068.  At a spatial scale of
0\farcs0287 per pixel these data provide an order of magnitude improvement
in resolution over previous ground based spectra and allow us to trace the
interaction between the radio jet and the gas in the NLR.
Our results show that, within $\pm 0\farcs5$  of the
radio-jet the emission lines are split into two components whose 
velocity separation is $\sim$ 1500 km \isec\ and are clearly
the result of the interaction between the radio jet and
the ambient NLR material.
The filaments associated with the radio lobe also show a redshifted kinematic disturbance of the order of 300 km \isec which probably is a consequence of the expansion of the radio plasma.

Our results show that the highest excitation 
gas is physically related to the location of the jet and is accompanied by an
excess continuum providing additional evidence for the interaction model.

The morphology, kinematics and, possibly, the ionization structure
of the NLR in the vicinity of the
jet of NGC 1068 are a direct consequence of the interaction with
the radio outflow.

\acknowledgements

A.M. acknowledges partial support through GO grant G005.44800 from
Space Telescope Science Institute, which is operated by the Association
of Universities for Research in Astronomy, Inc., under NASA contract
NAS 5--26555. A.C. thanks the STScI visitor program for providing financial support during the course of this work. A.R. is supported by a Royal Society Fellowship. We thank E. Oliva for kindly providing his
compilation of atomic parameters and his code to derive line
emissivities.

\clearpage

\end{document}